\documentclass[jcp,twocolumn,superscriptaddress]{revtex4}

\usepackage{epsfig}
\usepackage{graphics}
\usepackage{chemfig}
\usepackage{amsmath}

\begin{document}

%\twocolumn[\hsize\textwidth\columnwidth\hsize\csname
%@twocolumnfalse\endcsname

\title{A Chain-of-States Acceleration method for the efficient location of Minimum Energy Paths}

\author{E.~R.~Hern\'{a}ndez}
\thanks{email: Eduardo.Hernandez@csic.es} 
\affiliation{Instituto de Ciencia de Materiales de Madrid (ICMM--CSIC), Campus de Cantoblanco,
28049 Madrid, Spain}
\author{C.~P.~Herrero}
\affiliation{Instituto de Ciencia de Materiales de Madrid (ICMM--CSIC), Campus de Cantoblanco,
28049 Madrid, Spain}
\author{J.~M.~Soler}
\affiliation{Dep. F\'{\i}sica de la Materia Condensada and IFIMAC,  Universidad Aut\'{o}noma de Madrid, 28049 Madrid, Spain}

\date{\today}

\begin{abstract}
We describe a robust and efficient chain-of-states method for computing Minimum Energy Paths~(MEPs) associated to barrier-crossing events in poly-atomic systems, which we call the {\em Acceleration method\/}. The path is parametrized in terms of a continuous variable $t \in [0,1]$ that plays the role of time. In contrast to previous chain-of-states algorithms such as the Nudged Elastic Band or String methods, where the positions of the states in the chain are taken as variational parameters in the search for the MEP, our strategy is to formulate the problem in terms of the second derivatives of the coordinates with respect to $t$, {\em i.e.\/} the state {\em accelerations\/}. We show this to result in a very simple and efficient method for determining the MEP. We describe the application of the method to a series of test cases, including two low-dimensional problems and the Stone-Wales transformation in $\mbox{C}_{60}$.
\end{abstract}

% \pacs{62.50.-p, 64.70.dj, 65.20.-w, 65.40.-b}

\maketitle

\section{Introduction}
\label{sec:introduction}

In the study of molecular, condensed matter or materials systems  one is frequently confronted with the need to define a transition path for a given atomic rearrangement or chemical reaction. This involves specifying a curve in configuration space that goes from an initial state of local minimum energy, ${\bf r}_A$  (reactants), to a final one, ${\bf r}_B$, also of local minimum energy (products), that is representative of the manifold of actual trajectories through which the system could undergo the transition~\cite{Jonsson:1998p1578}. The most obvious and natural way to define such a curve is as a {\em Minimum Energy Path\/}~(MEP), i.e. a path that fulfills the condition of being a minimum of the potential energy surface~(PES) in the plane perpendicular to the path at any point along its length. Equivalently, the MEP is tangent to the PES gradient, and goes through at least one saddle point on its way from ${\bf r}_A$ to ${\bf r}_B$. 

There are several reasons why the MEP is a useful concept: firstly, as explained above, it gives a clear mathematical definition to the intuitive idea of {\em reaction mechanism\/}. The MEP allows to identify the energy barrier(s) and possible intermediate states of the transition in question. In systems where those barriers are significant (as compared to $k_B T$, frequently the case when the transition involves the breaking and forming of chemical bonds), identifying the relevant MEPs is a pre-requisite to the application of Transition-State Theory-based approaches to estimate the reaction rate. There are of course situations in which the MEP is not such a useful concept. This happens when there are many competing paths, none of them being overwhelmingly dominant~\cite{Chandler:1998}, which is the typical case in soft-condensed matter systems. Path-sampling techniques have been developed to estimate transition rate constants specifically in this kind of system~\cite{Bolhuis:2002p1668,Dellago:1998p1669,Sevick:1993p1548,Wales:2002p1616}. Nevertheless, in hard condensed matter and molecular systems, the norm is to have transitions that closely follow a well-defined path. This happens e.g. in the diffusion of atoms and defects in solids, either in the bulk or at surfaces, in many isomerization reactions in molecules, etc. Given the interest in this type of processes, it is hardly surprising that many algorithms devoted to finding MEPs have been developed (see~\cite{Jonsson:1998p1578} and references therein), and many practical applications of such algorithms have been reported in the literature. 

There are two strategies that have been frequently employed in order to identify and locate a MEP. The first one starts by locating the first-order stationary point (saddle point) that marks the position of the barrier between the two minima that one wishes to connect through the reaction path. This can be done in a number of ways, e.g. using a Hessian mode-following algorithm~\cite{Banerjee:85,Baker:1886p1649},  hybrid eigenvector following~\cite{Munro:1999p1643,Kumeda:2001p1683,Zeng:2014p1681}, or the {\em climbing-dimer\/} method~\cite{Henkelman:1999p1654}. Once the saddle point has been located, the MEP can be obtained by following the steepest descent path on either side of the barrier down to the relevant minima~\cite{Page:1988p1650}. The second strategy, and the one with which we will concern ourselves here, attempts to directly obtain the full path, usually represented as a {\em string of beads\/} or {\em state polymer\/}, in which each bead represents a configuration of the entire system displaced along the path. Methods of this kind are frequently referred to as {\em chain-of-states\/} methods, and some important examples are the {\em Nudged Elastic Band\/} (NEB)~\cite{Jonsson:1998p1578,Henkelman:2000p1539,Henkelman:2000p1655} and its variant, the {\em Doubly-nudged Elastic Band\/}~(DNEB)~\cite{Trygubenko:2004p1678}, the {\em String\/}~\cite{E:2002p1547,E:2007p1546} and the {\em Freely-jointed Chain\/} (FJC)~\cite{Sevick:1993p1548} methods, although there are others (for a review of earlier methods of this kind see~\cite{Jonsson:1998p1578}). The objective of this family of methods is to define a procedure that will cause the state polymer to evolve towards the MEP. Not only must the converged path fulfill (to within a specified numerical accuracy) the conditions for being an MEP; it must also retain the states evenly spaced along the chain in order to adequately discretize the MEP over its whole length. The NEB method achieves this by introducing harmonic spring potentials that couple each bead to its two nearest neighbors along the chain. The configuration of the state polymer is then updated by making each bead follow a direction in configuration space that is given by the composition of two forces:
\begin{equation}
{\bf f}_{i}^ {NEB} =  {\bf f}_i^{\perp} + {\bf f}_{i}^{spr\parallel}.
\label{eq:NEB}
\end{equation}
Here ${\bf f}_i^{\perp} = -\nabla E^{\perp}({\bf r}_i)$ is the force derived from the PES in state {\em i\/} projected onto the hyper-plane perpendicular to the path; this term tends to drive the configuration of the chain towards the MEP. The second term, ${\bf f}_{i}^{spr\parallel}$, is obtained from the force due to the harmonic springs, projected onto the local path tangent. The effect of this term is to keep the beads evenly spaced over the length of the path. In its original formulation~\cite{E:2002p1547}, the String method also uses the first term in Eq.~(\ref{eq:NEB}) to drive the state polymer towards the MEP; in contrast, this method does not use harmonic springs to keep the states evenly spaced, but rather uses an interpolation scheme (typically cubic spline interpolation~\cite{E:2007p1546}) to parametrize the path and re-distribute the beads at regular intervals along its length. In its more recent, simplified version~\cite{E:2007p1546}, the full PES force is used, as opposed to its path-normal projection, to drive the path towards the MEP. Finally, the FJC method uses a transformation from Cartesian to hyper-spherical coordinates, effectively imposing an even bead separation along the chain. Rather than evolving the chain in the direction of the normal force along each bead, this method minimizes the mis-alignment between the force on each bead and the local tangent. 

The NEB and String methods have been very successful, with numerous applications demonstrating their ability to locate MEPs in complex multi-dimensional systems. Although there are differences between the two (for recent studies comparing them see~\cite{Koslover:2007p1682,Sheppard:2008p1538}), they are very similar in spirit, with a common denominator being the fact that the chain-of-states configuration is evolved towards the MEP directly in configuration (coordinate) space. This is actually a feature that all chain-of-state methods that we are aware of have in common. In this work we contend that there is an alternative formulation of the problem in terms of the {\em acceleration\/} variables, resulting in a very simple algorithm that does not require the introduction of spring potentials or otherwise re-positioning beads along the chain to ensure an even discretization of the path. We term this algorithm the {\em Acceleration method\/}.

The structure of this paper is as follows: in Sec.~\ref{sec:methodology} we describe our formalism and strategy for locating MEPs. In Sec.~\ref{sec:results} we apply the method to a number of test cases, namely two simple toy models of reduced dimensionality, and a more realistic multi-dimensional problem involving an isomerization reaction in $\mbox{C}_{60}$. Finally, in Sec.~\ref{sec:conclusions} we review the main features of our method, point out some directions for future work and present our main conclusions.

\section{Methodology}
\label{sec:methodology}

Our starting point is a parametrization of the path between two stationary points on the PES. We will represent the path as follows:
\begin{equation}
{\bf r}(t) = (1-t)\, {\bf r}_{A} + t\, {\bf r}_{B} + {\bf u}(t).
\label{eq:path}
\end{equation}
Here ${\bf r}(t)$ is a vector of length $d \times N_{at}$, with $d$ being the space dimensionality (2 or 3 in the examples discussed in Sec.~\ref{sec:results}) and $N_{at}$ the number of atoms in the system; $t \in [0,1]$ is a reaction parameter, such that ${\bf r}(0) = {\bf r}_A, {\bf r}(1) = {\bf r}_B$, with ${\bf r}_A$ and ${\bf r}_B$ being the given start and end configurations, which are stationary points (typically minima) of the PES on which we seek to find an MEP; and ${\bf u}(t)$ measures the deviation of the path from the linear interpolation and, by construction, must fulfill the boundary conditions ${\bf u}(0) = {\bf u}(1) = {\bf 0}$. Another requirement we impose on ${\bf u}(t)$ is that its components be continuous and twice-differentiable functions of $t$. The objective, then, is to find ${\bf u}(t)$ for~$t \in [0,1]$ such that Eq.~(\ref{eq:path}) is an MEP of the PES between ${\bf r}_A$ and ${\bf r}_B$. This will happen when the gradient of the PES at any point $t$, $\nabla E[{\bf r}(t)]$, is co-linear to the path (whenever $\nabla E[{\bf r}(t)] \neq {\bf 0}$), or, in other words, when the gradient component perpendicular to the path is zero. 
Because Eq.~(\ref{eq:path}) constitutes an analytical representation of the path, for any given trial path we can calculate the path tangent, ${\bf v}(t) = d{\bf r}(t)/d t$, i.e. the {\em velocity\/}, if we view variable $t$ as (fictitious) {\em time\/}. Likewise, we can also calculate the {\em acceleration\/}, ${\bf a}(t) = d^2{\bf r}(t)/d t^2$. In particular, ${\bf v}(t)$ is important, since it provides us with a criterium for MEP convergence (${\bf v}(t)$ and $\nabla E[{\bf r}(t)]$ must be co-linear). As we shall see below, ${\bf a}(t)$ also plays a major role in our scheme. Notice that, given the two boundary conditions, there is a biunivocal relationship between  ${\bf u}(t)$ and ${\bf a}(t)$.

Although Eq.~(\ref{eq:path}) offers a continuum representation of the path, in practical calculations it is necessary to resort to a discrete representation in terms of a set of $N$ replicas of the system, ${\bf r}(t_n)$, where $t_n = n \, \Delta t,\; \Delta t = 1/(N-1), \; n = 0, 1, \ldots N-1$. This amounts to specifying the components of the ${\bf u}$ vector at $N$ points (including the end points) along the path. Discretizing the path in this way does not really pose a drawback, as it is always possible to effectively recover an analytical representation by means of an interpolation scheme, or by using a set of suitable continuous functions of $t$ to expand the components of ${\bf u}$. This can always be done provided $N$ is not too small.

Let us now consider the problem of varying the path in search of an MEP. As noted in Sec~\ref{sec:introduction} above, previous chain-of-states methods use the coordinates of the beads along the path as variational parameters in the MEP search. As it is well-known~\cite{Jonsson:1998p1578}, directly optimizing a path in terms of  bead coordinates will result in a highly winding path with unevenly distributed beads, and in general does not converge towards an MEP. Different strategies can be adopted to avoid this problem (harmonic springs coupling neighboring beads in the NEB method, reparametrizing the path at regular intervals, as in the String method, etc). 
In this work, however, we argue that the practical difficulties arising from using the coordinates as variational parameters can be very naturally overcome by using instead the accelerations, ${\bf a}_n \equiv {\bf a}(t_n)$, as variational degrees of freedom. The idea is simply to adjust iteratively ${\bf a}_n$ in order to drive the path towards a configuration fulfilling the requirement that the force perpendicular to the path, ${\bf f}_n^\perp = -\nabla E^{\perp}[{\bf r}(t_n)] = {\bf 0}$ for~$n \in [0,N-1]$. Our method is summarized as follows: 
\begin{enumerate}
\item Given an initial configuration of the path (e.g. the linear interpolation between start and end configurations, although other choices are possible) and its discretization by means of a number $N$ of replicas, construct its representation via Eq.~(\ref{eq:path}) using some appropriate interpolation scheme to define the ${\bf u}(t)$ functions (see below). From this representation calculate ${\bf v}_n \equiv {\bf v}(t_n)$ and ${\bf a}_n$ for every bead along the path.
\item Calculate the force at each bead position, and from it and the path tangent ${\bf v}_n$, obtain the force component perpendicular to the path, i.e. ${\bf f}_n^\perp= {\bf f}_n - ({\bf f}_n \cdot \hat{\bf v}_n)\, \hat{\bf v}_n$, where $\hat{\bf v}_n \equiv {\bf v}_n/|{\bf v}_n|$.
\item Update the acceleration vector according to: ${\bf a}_n \leftarrow {\bf a}_n - \lambda {\bf f}_n^\perp$, where $\lambda$ is a positive numerical parameter, having dimensions of inverse mass, to be suitably adjusted so as to optimize convergence towards the MEP. 
\item By integrating a suitable interpolation ${\bf a}(t)$ of the new ${\bf a}_n$, obtain new vectors ${\bf v}_n$ and ${\bf u}_n$. The integration constants are fixed by the boundary conditions ${\bf u}(0) = {\bf u}(1) = {\bf 0}$.
\item Return to step 2 above, and iterate the procedure until the path converges to the MEP. 
\end{enumerate}

Before discussing the details of our practical implementation of the above scheme the following comments are in order. Firstly, the need to perform a double integration in $t$ to obtain the new path configuration from ${\bf a}_n$ (step 4) may be perceived to be a disadvantage of the method. However this is not so: the path, and in particular the MEP, is generally a smooth, low-curvature trajectory in configuration space. It follows that the components of ${\bf u}(t)$ are also smooth, well-behaved and slowly-varying functions of $t$. Therefore, provided $N$ is not too small, and a decent interpolation scheme is used, it is possible to insure that the integration is performed with sufficient accuracy. A more fundamental reason to work in terms of accelerations is discussed at the end of this section. Secondly, by viewing the path as a trajectory, and $t$ as its time variable, it is easy to see that step 3 above changes only the path-normal component of the acceleration. This component affects only the shape of the path, i.e. the direction of its tangent vector, ${\bf v}(t)$, but not its modulus. It follows that images are not caused to slide up or down the path in any significant way, and thus the inter-bead spacing will (to first order) remain even. Nevertheless, inter-bead spacing will become uneven over a sufficiently large number of iterations of the scheme due to curvature effects. If needed, the tangential components of the acceleration can be scaled by a factor smaller than 1, so as to gradually reduce their value during the iterative process, which will ensure even spacing of the beads. In the examples that follow we found this to be unnecessary, although we did it in the second example for illustrative purposes. Thirdly, the parameter $\lambda$ introduced in step~3 above determines the rate of convergence of the method, and choosing it well is therefore important. In the illustrative examples discussed in Sec.~\ref{sec:results} we have for simplicity adopted the strategy of taking it as a constant value, giving overall adequate convergence. However better convergence rates may be achieved by allowing $\lambda$ to vary and using e.g. a Hessian-update scheme to choose $\lambda$ appropriately at each step and/or for each bead independently. This issue will be the subject of future research.

The general scheme described above could be implemented in a number of different ways; all that is needed is a flexible interpolation scheme that allows to construct a representation of the ${\bf u}(t)$ vector components from the bead positions, or rather, their second derivative with respect to $t$ (which enter in the acceleration), and to perform the reverse process of integrating ${\bf a}_{new}(t)$ to obtain the new configuration of the path (step 4). This could be done e.g. using cubic spline interpolation, or any other suitable interpolation scheme. 
%In particular, we have found a Fourier representation of the ${\bf u}$ vector components to be particularly convenient. In our implementation we represent them as follows:
%\begin{equation}
%u_{\alpha,n} = \frac{1}{N} \sum_{k=0}^{N-1} \tilde{u}_{\alpha,k}\; e^{-2 \pi i k n/N}.
%\label{eq:Fourier}
%\end{equation}
%Here $N$ is the number of images or replicas with which the path is discretized, and $u_{\alpha,n}$ is the $\alpha$-component of ${\bf u}(t)$ at point $t_n$. The $\tilde{u}_{\alpha,k}$ can be obtained from the $u_{\alpha,n}$ by a discrete Fourier transform. Obviously, since the ${\bf u}(t)$ vector has real components, only the real part of Eq.~(\ref{eq:Fourier}) need be considered. Eq.~(\ref{eq:Fourier}) allows for a straight-forward calculation of the first and second {\em t\/}-derivatives of $u_{\alpha,n}$ (needed in step 1 above) in Fourier space, which can then be transformed to real space by means of an inverse Fourier transform. Likewise, the double integration needed in step 4 to obtain the updated Cartesian configuration of the path is trivially performed in Fourier space. Because of the boundary conditions and the smoothness of  the ${\bf u}$ components, Eq.~(\ref{eq:Fourier}) converges rapidly with the number of replicas in the discretization. Another desirable feature of Eq.~(\ref{eq:Fourier}) is that it allows for a straight forward refinement of the path representation, should this be required, by means of Fourier interpolation~\cite{code}. 
In particular, we have found a Fourier sine series representation of the ${\bf u}$ vector components to be particularly convenient. In our implementation we represent them as follows:
\begin{equation}
{\bf u}(t) = \sum_{n=1}^{N-2} \tilde{{\bf u}}_n \sin( \omega_n t),
\label{eq:Fourier}
\end{equation}
where $\omega_n = n\pi$. The $N-2$ nonzero Fourier coefficients $\tilde{{\bf u}}_n$ are fixed by the $N-2$ nonzero values ${\bf u}_n$, with $0<n<N-1$. Eq.~(\ref{eq:Fourier}) obeys the boundary conditions ${\bf u}(0) = {\bf u}(1) = 0$ by construction. Another advantage is that the first and second derivatives ${\bf u}'(t)$, and  ${\bf u}''(t)$, are similarly given as cosine and sine series, respectively. It is therefore very simple to obtain ${\bf u}(t)$ and ${\bf v}(t)$ from ${\bf a}(t)$, as required by step~4 of our algorithm. Indeed, following step~3 one obtains new accelerations ${\bf a}(t) = {\bf u}''(t)$, which, by virtue of Eq.~(\ref{eq:Fourier}), have components of the form
\begin{equation}
{\bf a}(t) = \sum_{n=1}^{N-2} \tilde{{\bf a}}_n \sin( \omega_n t ),
\label{eq:FourierAcceleration}
\end{equation}
where again the Fourier coefficients $\tilde{\bf a}_{n}$ can be obtained from ${\bf a}_{n}$ by Fourier transform techniques. Now, Eq.~(\ref{eq:FourierAcceleration}) can be integrated two times to give
\begin{eqnarray}
{\bf v}(t) & = & -\sum_{n=1}^{N-2} \frac{\tilde{{\bf a}}_n}{\omega_n} \cos( \omega_n t) + {\bf C}_0, \\
\label{eq:FourierVelocity}
{\bf u}(t) & = & -\sum_{n=1}^{N-2} \frac{\tilde{{\bf a}}_n}{\omega_n^2} \sin( \omega_n t ) + {\bf C}_1.
\label{eq:FourierPosition}
\end{eqnarray}
The boundary conditions fix the values of the integration constants to be ${\bf C}_0={\bf r}_B-{\bf r}_A$ and ${\bf C}_1=0$. 

In the next section we will show that the method just described is robust, efficient and stable. Before we describe its application to specific examples, it is worth pausing to reflect on the reasons for its stability. One may naively assume that a similar scheme to ours, but formulated in terms of coordinates instead of accelerations (i.e. using ${\bf r}_n \leftarrow  {\bf r}_n + \lambda {\bf f}_n^\perp$ in step~3) should work just as well, thus obviating the need to integrate accelerations to obtain the coordinates and velocities. However practical experience shows that this is not the case, as is well documented~\cite{Jonsson:1998p1578}. Such a scheme results in a snake-like path dominated by high-frequency error components that never converges to the MEP. This, however, does not happen in our scheme, and in Eq.~(\ref{eq:FourierPosition}) we can see the reason for this: in the integration step to obtain the coordinates each acceleration component is scaled by the inverse of its corresponding frequency squared, thus effectively acting as a filter to high frequency error components. As a consequence, the path evolves more smoothly towards the MEP, which allows a faster convergence without developing kinks or twists in the process.

\section{Results}
\label{sec:results}

In order to demonstrate the effectiveness of the methodology presented in Sec.~\ref{sec:methodology} above, we describe here its performance in three specific cases of MEP location. The first two examples we consider are simple 2D potential energy surfaces (PES), namely the modified LEPS potential (model II in~\cite{Jonsson:1998p1578}) and the M\"{u}ller potential~\cite{Mueller:1979p1652}. As a third example we consider the multi-dimensional problem of the Stone-Wales isomerization transition~\cite{stone86} between the $I_h$ (buckminsterfullerene) and the $C_{2v}$ lowest energy isomers of $C_{60}$. The first two examples are simple toy models of reduced dimensionality, but nevertheless contain all the essential ingredients of the problem in the more general, multi-dimensional case. In spite of their simplicity and 2D character, they constitute challenging test cases for any methodology that aims to be a viable alternative for the location of MEPs. In all cases discussed below we took as initial guess for the MEP a simple linear interpolation between the end points of the path, which invariably were chosen as two previously located minima on the corresponding PES.  The number of beads or replicas of the system along the path was varied between a minimum of 10 and a maximum of 30, although individual tests have been also made with bead numbers outside this range. 

\subsection{Modified LEPS potential}

\begin{figure}[!t]
\centering
\begin{minipage}[t]{8cm}
\epsfxsize=8cm
\epsffile{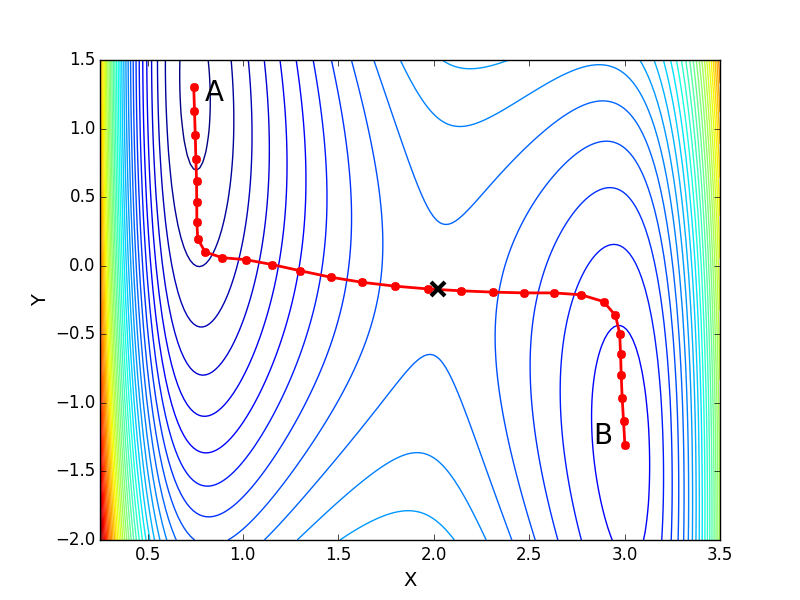}
\end{minipage}
\begin{minipage}[t]{8cm}
\epsfxsize=8cm
\epsffile{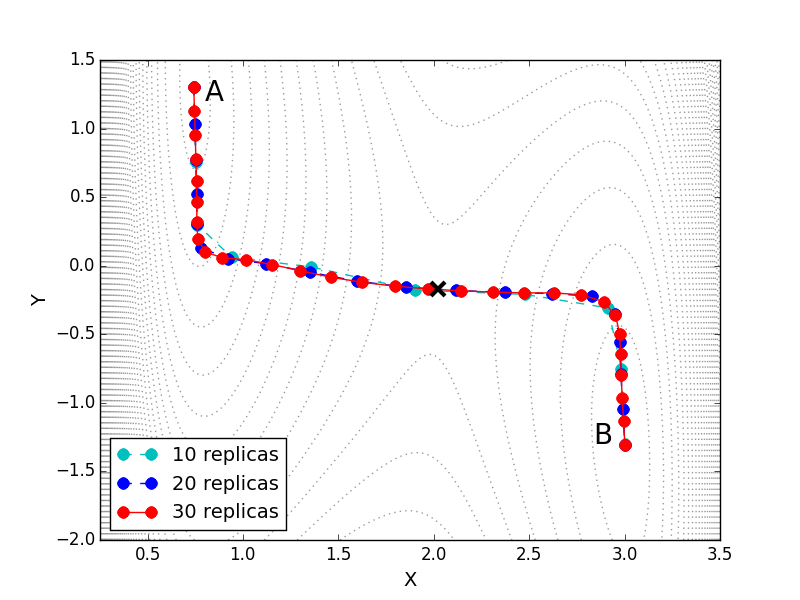}
\end{minipage}
\caption{(Color online) The top panel shows the MEP obtained for the LEPS potential using 30 replicas of the system along the path. The black cross marks the position of the saddle point between the two minima, and the minima, marking the start and end of the path, are labeled as A and B. For comparison, the lower panel displays simultaneously three MEPs obtained with 10, 20 and 30 beads.
}
\label{fig:LEPSMEP}
\end{figure} 

The modified LEPS potential (see~\cite{Jonsson:1998p1578} for details of its definition) possesses two minima, the first one of which is located at $x = 0.7415, y = 1.3034$, will be labeled as A in what follows, and is the global minimum on this PES. The second, local minimum B, is found at $x=3.0012, y = -1.3040$. A barrier separates the valleys of each minimum, with a saddle point located at $x = 2.021, y = -0.173$. A contour plot of this potential is shown in Fig.~(\ref{fig:LEPSMEP}). Also shown in the figure is the converged MEP~\cite{code} that resulted with 30 beads in the path discretization. As expected, the obtained MEP cuts perpendicularly the PES contour lines, and passes through the saddle point at the top of the barrier between the A and B valleys (marked with a black cross on the figure). Note how the replicas (shown as red dots) remain roughly evenly spaced along the MEP, in spite of the fact that no harmonic springs are used to impose this, in contrast to the case of the NEB method; neither have they been artificially redistributed, as in the String method. Even the top of the barrier remains well described by a sufficient density of beads. 

\begin{figure}[!t]
\begin{center}
\epsfxsize=8cm
\epsffile{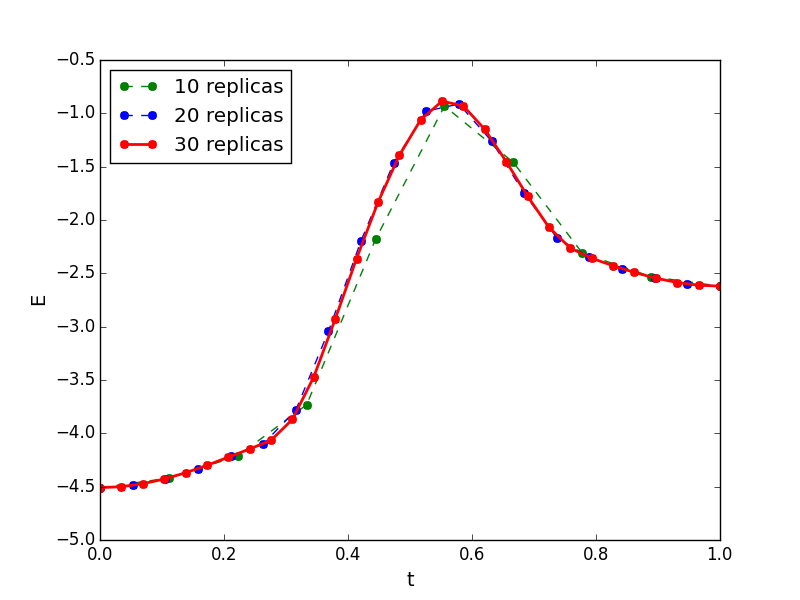}
\end{center}
\caption{(Color online) The energy profile of the LEPS potential along the MEP obtained with 10, 20 and 30 replicas of the system along the path.
$t = 0$ corresponds to the first minimum [A, see Fig.~(\ref{fig:LEPSMEP})], and $t = 1$ to the second one (B).
}
\label{fig:LEPSMEPProfile}
\end{figure}

It is worth noticing that in spite of the simplicity and reduced dimensionality of this PES, the MEP has sharp bends, where the path turns by nearly 90 degrees (when climbing out of the A valley and down into the B valley). Such regions of high curvature would pose a challenge for any simplistic approach to MEP location, but our methodology encounters no particular difficulty with these regions.

Fig.~(\ref{fig:LEPSMEPProfile}) displays the energy profiles along the MEP when the latter is discretized with 10, 20 and 30 beads. As can be seen there, using only 10 beads results in a relatively rugged description of the MEP, although the general features of the path, such as the barrier height, are reasonably well reproduced even in this case. With 20 and 30 beads a much smoother and accurate representation of the MEP is obtained, as evidenced from the fact that the energy profiles in these cases are hardly distinguishable on the plot. The description of the barrier summit is slightly more accurate with 30 replicas due to the increased density of beads, but elsewhere the two profiles are practically identical. 

\subsection{M\"{u}ller potential}

\begin{figure}[!t]
\centering
\begin{minipage}[t]{8cm}
\epsfxsize=8cm
\epsffile{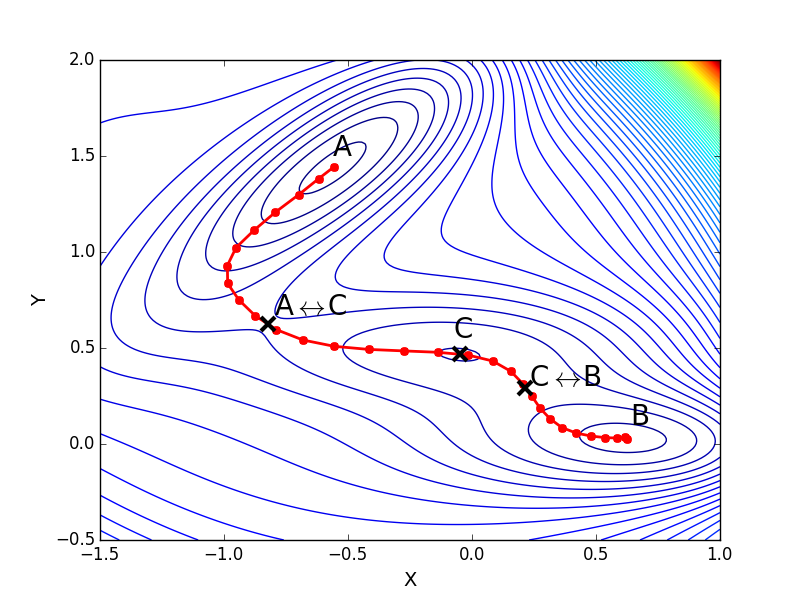}
\end{minipage}
\begin{minipage}[t]{8cm}
\epsfxsize=8cm
\epsffile{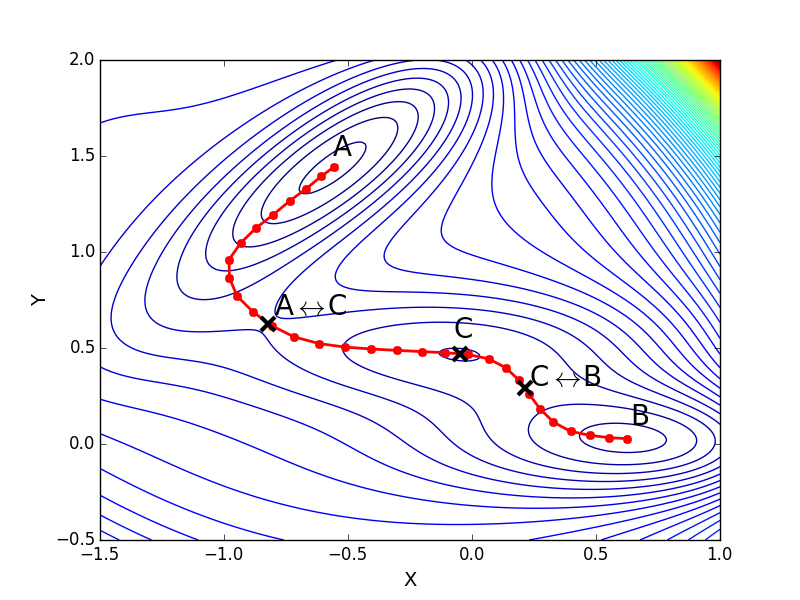}
\end{minipage}
\caption{(Color online) The MEP obtained for the M\"{u}ller potential using 30 replicas of the system along the path. The black crosses mark the position of an intermediate local minimum and of two saddle points between the two end minima. The minima  are labeled as A (the absolute minimum),  B,  and the intermediate one, C. Two saddle points mark the position of the barriers separating the minima,  the first one being labeled as A$\leftrightarrow$C, and the second one labeled as C$\leftrightarrow$B. The upper panel shows the MEP obtained without performing any scaling of the tangential acceleration components; as can be seen, the bead distribution becomes somewhat uneven. In the lower panel, scaling the tangential acceleration components by a factor of 0.99 at every convergence iteration results in an even distribution of beads along the MEP. 
}
\label{fig:MullerMEP}
\end{figure} 

Let us now consider the case of the M\"{u}ller potential~\cite{Mueller:1979p1652}. In contrast to the  LEPS model seen above, this PES has three minima, and two saddle points separating them. Although still only a 2D model, the presence of more stationary points on the PES constitutes an added challenge for MEP location algorithms. A contour plot of this PES is shown in Fig.~(\ref{fig:MullerMEP}), together with the location of the various stationary points. The minima are labeled as A (the absolute minimum), with coordinates $x=-0.558, y = 1.442$,  B, $x=0.623, y = 0.028$, and the intermediate one, C, located at $x = -0.05, y = 0.467$. Two saddle points separate the valleys corresponding to each minimum, at $x=-0.822, y=0.624$ (labeled as A$\leftrightarrow$C) and $x=0.212, y=0.293$ (labeled as C$\leftrightarrow$B), respectively. Like in the case of the LEPS model, we considered path discretizations using 10, 20 and 30 beads, and in each case the initial path was taken as the linear interpolation between A and B. In the upper panel of Fig.~(\ref{fig:MullerMEP}) we plot the resulting MEP~\cite{code} using 30 replicas, without using any scaling of the tangential acceleration components to keep the beads evenly spaced. As expected, the obtained MEP goes through the intermediate minimum (C) and the two saddle points located along the path. While the finest description of the MEP is obtained with 30 replicas, coarser path descriptions with 10 and 20 beads (not shown) also track the MEP correctly. The converged MEPs have beads roughly evenly distributed along the whole length, even without scaling the tangential components of the acceleration. For comparison, in the lower panel of Fig.~(\ref{fig:MullerMEP}) we show the MEP obtained when the tangential acceleration components are scaled by a factor 0.99 at each step of the iterative process; as can be seen, the MEP that results in this case has homogeneously distributed beads. 

\begin{figure}[!t]
\begin{center}
\epsfxsize=8cm
\epsffile{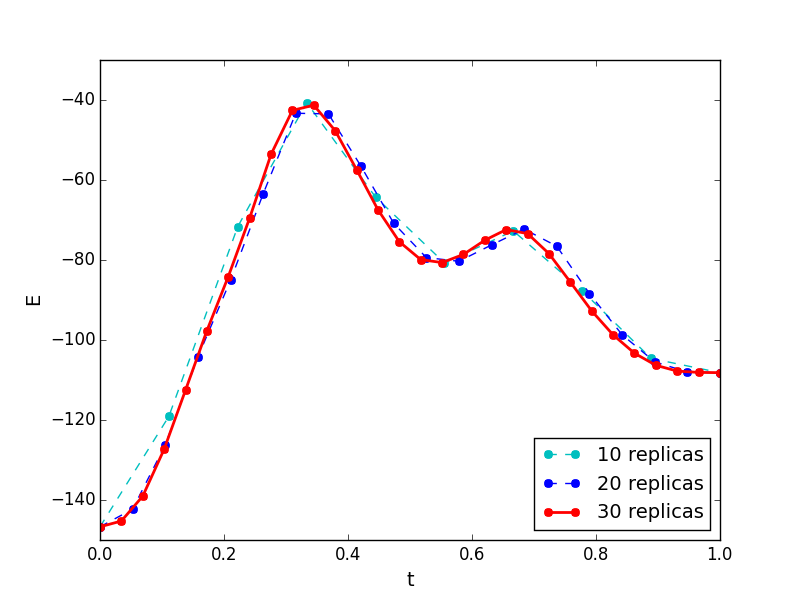}
\end{center}
\caption{(Color online) The energy profile of the M\"{u}ller potential along the MEP obtained with 10, 20 and 30 replicas of the system along the path.
$t = 0$ corresponds to the first minimum [A, see Fig.~(\ref{fig:MullerMEP})], and $t = 1$ to the second one (B).
}
\label{fig:MullerMEPProfile}
\end{figure}

Fig.~(\ref{fig:MullerMEPProfile}) shows the energy profile along the converged MEPs with the different path discretization used in the calculation. As can be seen by comparing Figs.~(\ref{fig:MullerMEPProfile}) and~(\ref{fig:LEPSMEPProfile}), the energy profile on the M\"{u}ller PES has more features (two peaks and an intermediate valley) than that of the LEPS model. Using only10 beads to describe the MEP results in a rather coarse description of the path, but even at this level all features of the energy profile are captured and reasonably described. 

\subsection{Stone-Wales transformation in $\mbox{C}_{60}$}

\begin{figure}[!t]
%\schemestart
%\setbondstyle{line width = 2pt}
%\chemfig{*6(-(-[:-60]-[:0,0.75]-[:60])-[,,,,dash pattern = on 2pt off 2pt]*6(----(-[:120]-[:180,0.75]-[:-120])-[,,,,dash pattern = on 2pt off 2pt]-)-[,,,,red]---)}
%\arrow{<=>} 
%\setbondstyle{line width = 2pt}
%\chemfig{[:-30]*6(---(-[:30]-[:90,0.75]-[:150])-*6(-[,,,,dash pattern = on 2pt off 2pt]---(-[:210]-[:270,0.75]-[:-30])--)-[,,,,red]-[,,,,dash pattern = on 2pt off 2pt])}
%\schemestop
\epsfxsize=8cm
\epsffile{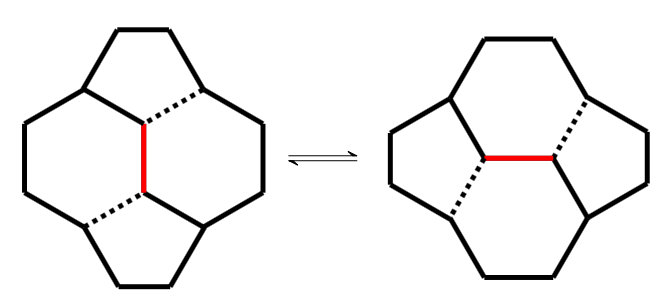}
\caption{(Color online) Scheme of the Stone-Wales transition. The central bond, highlighted in red, rotates by $90^\circ$; 
for this to happen two bonds have to be broken,
such as the ones marked by dashed lines, and re-formed after the rotation of the central bond takes place. }
\label{fig:Stone-Wales}
\end{figure}

As a final example we will consider the case of the Stone-Wales~(SW) transformation~\cite{stone86} in $\mbox{C}_{60}$. It is known that $\mbox{C}_{60}$ has 1812 fullerene isomers (i.e. cage structures formed exclusively by 20 hexagons and 12 pentagons). Only one of these isomers obeys the {\em isolated pentagon rule\/}~\cite{Kroto:1987p1670}; it is the $I_h$ structure known as buckminsterfullerene, having every pentagon surrounded by five hexagons (there are no pentagon-pentagon adjacencies in this structure). This is the most stable structure of $\mbox{C}_{60}$; in all other $\mbox{C}_{60}$ isomers pentagon-pentagon adjacencies are present, incurring an energy penalty that renders these structures less stable than the $I_h$ isomer. Stone and Wales~\cite{stone86} were the first to point out that it was possible to transform a given fullerene isomer into a different one by  means of the rotation of a C-C bond connecting two hexagons and two pentagons. The rotation of such a bond around its center interchanges the positions of the hexagons and pentagons, as illustrated in Fig.~(\ref{fig:Stone-Wales}). Given the importance of the SW transformation in the growth of carbon nanostructures~\cite{hernandez:2001} and as a stress-release mechanism in carbon nanotubes~\cite{buongiorno98a,buongiorno98b}, it has been extensively studied at different levels of theory (see e.g. ref.~\cite{Kumeda:2003p1684,Bettinger:1915p1573} and references therein).

\begin{figure}[!t]
\begin{minipage}[t]{2.5cm}
\epsfxsize=2.5cm
\epsffile{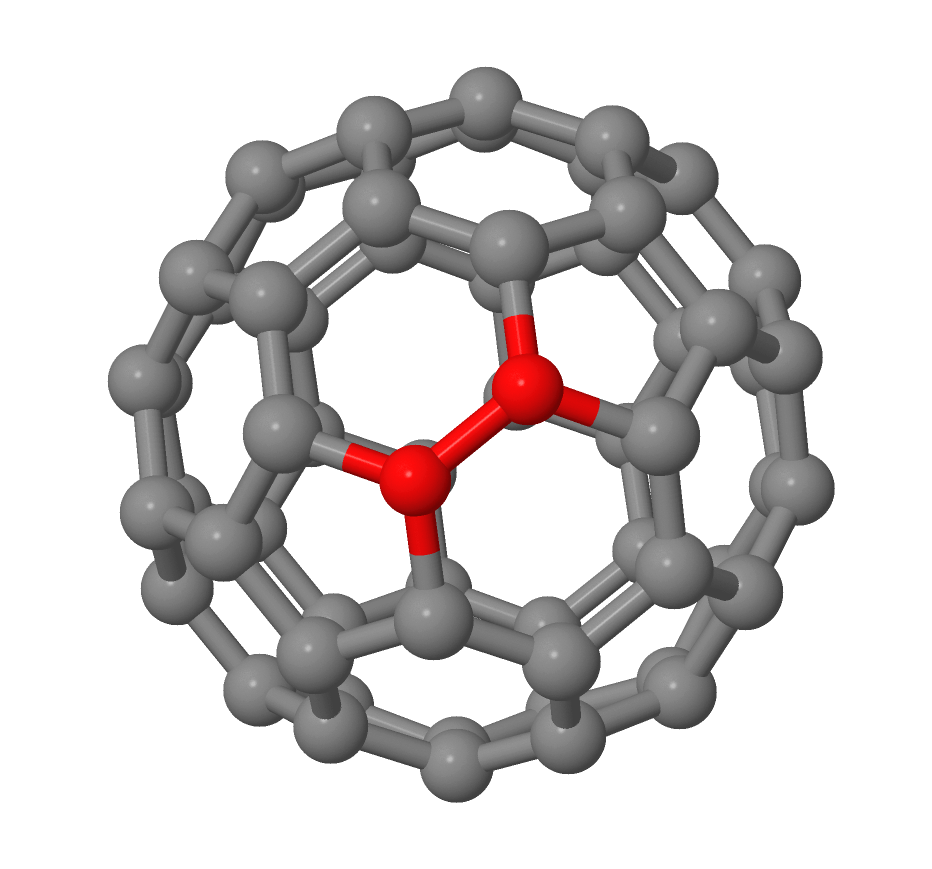}
\end{minipage}
\begin{minipage}[t]{2.5cm}
\epsfxsize=2.5cm
\epsffile{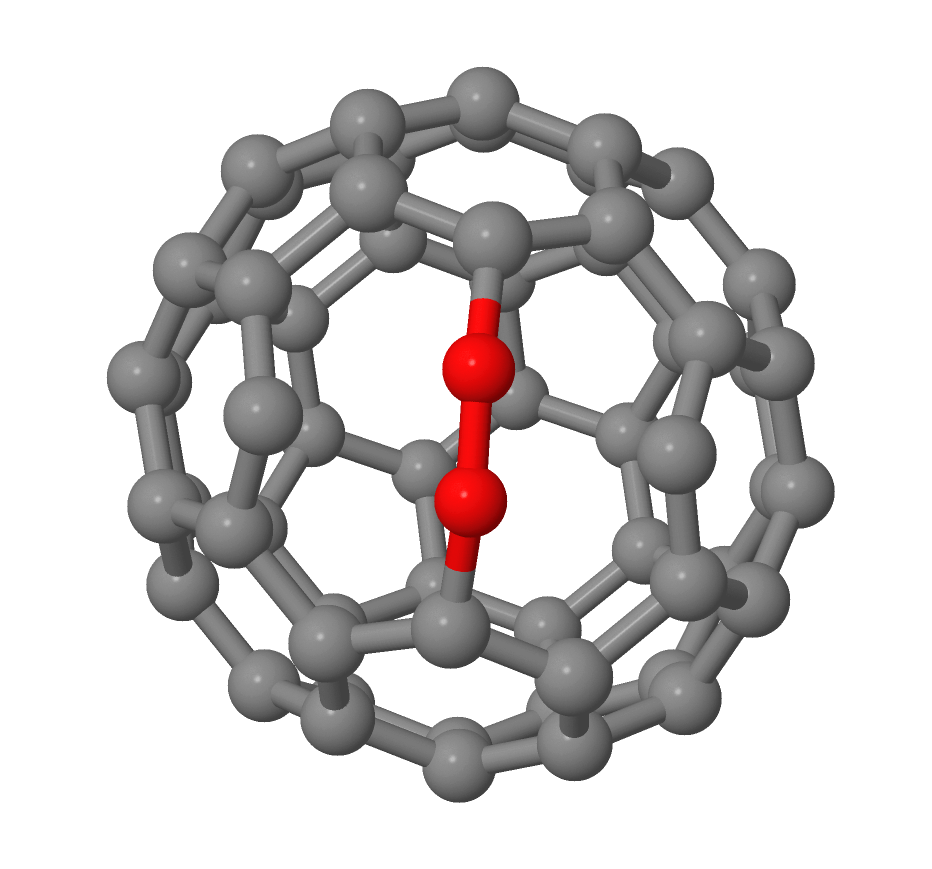}
\end{minipage}
\begin{minipage}[t]{2.5cm}
\epsfxsize=2.5cm
\epsffile{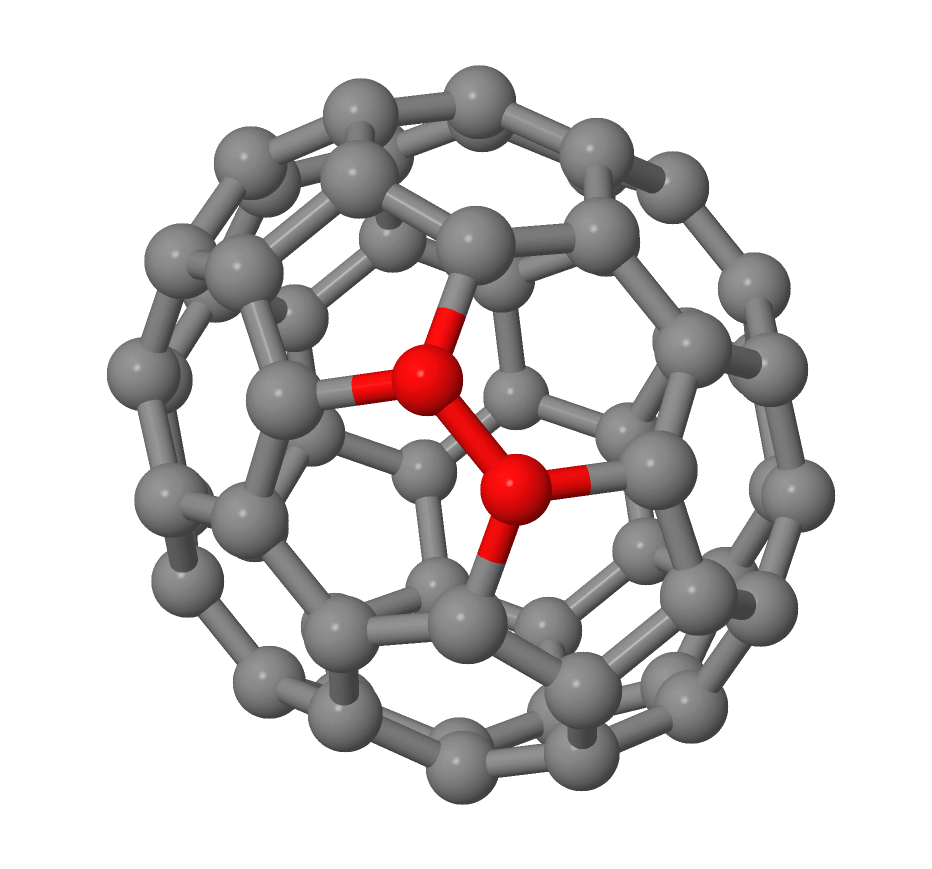}
\end{minipage}
\begin{minipage}[b]{8cm}
\epsfxsize=8cm
\epsffile{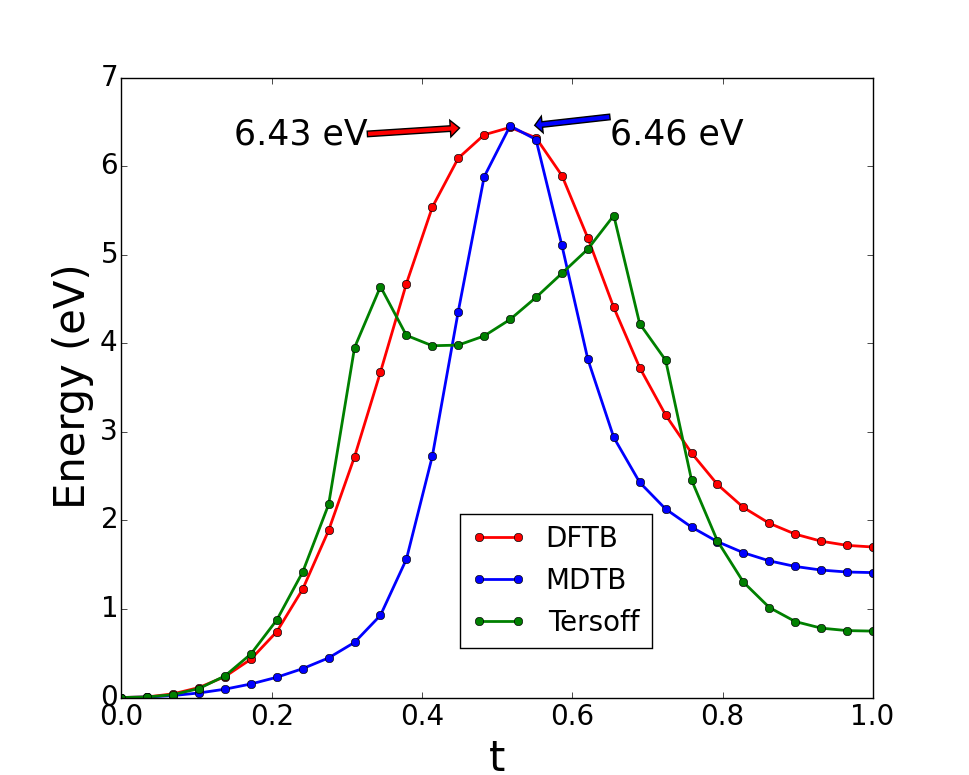}
\end{minipage}
\caption{(Color online) The energy profile along the Stone--Wales transition MEPs in $\mbox{C}_{60}$. The structures shown correspond to the $I_h$ isomer (buckminsterfullerene, the ground state) on the left, the $C_{2v}$ isomer on the right, and the saddle point configuration between the two, as calculated with the DFTB model. MEPs  obtained with the Tersoff potential and the MDTB and DFTB tight binding models are shown for comparison. Red and blue arrows mark the energy of the DFTB and MDTB saddle points, respectively (see text). }
\label{fig:C60MEP}
\end{figure}

Here we will consider the SW transformation between the $I_h$ buckminsterfullerene and the $C_{2v}$ lowest-energy isomers of $\mbox{C}_{60}$. 
The energetics of the system has been described with three different models, namely the many-body potential due to Tersoff~\cite{tersoff88}, the orthogonal tight-binding~(TB) model of  Xu {\em et al.\/}~\cite{Xu:1992p1639}, known as MDTB, and the non-orthogonal TB model model due to Porezag {\em et al.\/}~\cite{DFTB}, known as DFTB. All these models have been extensively used in the study of carbon-based systems. To perform the calculations described below we have coupled the MEP-search method described in Sec.~\ref{sec:methodology} with the Trocadero code~\cite{trocadero}, which contains implementations of all these energy models. The connecting path between the $I_h$ and $C_{2v}$ isomers has been discretized with 30 replicas; tests with different numbers of replicas (both smaller and larger than 30) were also carried out, leading to practically indistinguishable MEPs. 
%The energetics of the system have been described with the total-energy tight-binding~(TB) model due to Porezag {\em et al.\/}~\cite{DFTB}. While only semi-empirical in nature, this model has proved to be very successful in its description of carbon-based nanostructures. The connecting path between the $I_h$ and $C_{2v}$ isomers was discretized with 20 and 30 images in two different calculations, but both descriptions resulted in practically indistinguishable MEPs. 
In Fig.~(\ref{fig:C60MEP}) we plot the energy profiles along the MEP that resulted from these calculations. The energy of the ground state (isomer $I_h$) is taken as the zero of energy. Let us first discuss the MEP that results with the Tersoff potential. According to this model, the energy difference between the $I_h$ and $C_{2v}$ isomers is only 0.75~eV, much below typical values found with first-principles electronic structure methods, which are about twice as large~\cite{Kumeda:2003p1684,Bettinger:1915p1573}. Furthermore, the MEP obtained from the Tersoff description is very unusual, showing two sharp peaks with discontinuous derivative either side of a local minimum at the center of the barrier. The peaks do not become smoother if a larger number of beads is used to discretize the path; they appear to be features of this model potential. %which is known to have force discontinuities due to its functional form. 
The local minimum found close to the center of the barrier appears because this potential tends to over-stabilize the carbon atoms at the edge of the SW motif in the central configuration displayed at the top of Fig.~(\ref{fig:C60MEP}), turning the structure into a local minimum, instead of the saddle point that one would expect to find close to the barrier center. The maximum barrier height found along the Tersoff MEP is 5.44~eV, somewhat lower than that found by Marcos {\em et al.\/}~\cite{Marcos:1997p1661} (5.58~eV) using the same model; these authors reported a low-energy path from the $I_h$ to $C_{2v}$ isomers by linear interpolation between a series of intermediate minima that they identified from molecular dynamics simulations. The overall shape of their energy profile is similar to the MEP reported here, but contains actually three local minima, not counting the end structures. We believe that the two local minima on either side of the barrier they report are actually not there, and are seen in their energy profile because their path is not a true MEP. Marcos {\em et al.\/}~\cite{Marcos:1997p1661} attached some physical significance to the local minima they found, arguing that they would facilitate the global SW transition, as each intermediate step is subject to lower barriers than the overall process. We rather believe that the local minimum at the center of the barrier (the only one we find) is an artifact that results from the poor transferability of the Tersoff potential to situations far departed from the configurations considered in its parameter fitting. 

The main difference between the Tersoff potential and the TB models discussed next is that the latter incorporate a description of the (valence) electronic structure of the carbon atoms, albeit at a semi-empirical level~\cite{goringe97}. The TB models themselves are similar in spirit, but differ mainly in the fact that the MDTB model assumes the underlying basis set to be orthonormal, while the DFTB model explicitly incorporates their overlap, which in principle makes the model more transferable. These TB models provide a much more credible picture of the SW transition, that is in better agreement with what is predicted by higher levels of theory. Indeed, the MDTB model predicts the energy difference between the end isomers to be 1.41~eV, while the DFTB model gives a value of 1.7~eV, both much closer to the range of values resulting from first-principles calculations, which average at about 1.8~eV~\cite{Kumeda:2003p1684,Bettinger:1915p1573}. Both models also provide similar barrier heights: 6.45~eV (MDTB) and 6.44~eV (DFTB), which again are close to values predicted by higher levels of theory, averaging at about 7.5~eV. As can be seen in Fig.~(\ref{fig:C60MEP}), none of the TB models predicts the existence of any local minimum along the MEP, in contrast to the Tersoff potential. The obtained paths are smooth, without sharp features. 
%According to the TB model employed here, the energy difference between the$C_{2v}$ and $I_h$ isomers is 1.7~eV, and the barrier height, determined from the highest energy found along the MEP, is found to be 6.5~eV. These values fall within the ranges predicted by previous work using a variety of different levels of theory, including semi-empirical and first-principles electronic structure methods. For the $I_h-C_{2v}$ energy difference the predicted values  range from 1.47 to 3.2~eV, with the majority falling towards 1.8~eV; for the barrier height the range of values is within 6.2 and 8.5~eV, with an average around 7.5~eV (see~\cite{Bettinger:1915p1573} and references therein). 
In spite of the higher dimensionality of this problem ($3N-6=174$ degrees of freedom),  compared to the LEPS and M\"{u}ller potentials considered above, it is still the case that the replicas remain roughly evenly spaced along the converged MEPs, even in the case of the Tersoff potential description, in the presence of the sharp features observed there.

The red and blue arrows on either side of the barriers shown in the main panel in Fig.~(\ref{fig:C60MEP}) mark the energy of the transition-state configuration according to the DFTB and MDTB models, respectively. These configurations have been located using the mode-following Rational Function Optimization~(RFO) approach of Banerjee {\em et al.\/}~\cite{Banerjee:85,Baker:1886p1649}. This algorithm searches for stationary points of the PES using both first-derivative (gradient) and of second-derivative (Hessian) information. A saddle point can be located following a chosen eigenvector of the Hessian up-hill in energy, while minimizing along all the other modes. In our case we maximized along the eigenvector associated to the lowest Hessian eigenvalue, starting from the highest energy configuration found along the MEP. The search for the saddle point was assumed to have converged once the resulting structure had a gradient with components smaller than $10^{-5}$~eV/\AA\ and the Hessian had a single negative eigenvalue. The final DFTB configuration is illustrated as the central structure at the top of Fig.~(\ref{fig:C60MEP}); the one obtained with the MDTB model is very similar and is not shown. As can be seen in the figure, the energies of the transition states located in this way are very close to the barrier heights predicted from the MEPs, indicating that these are well converged.

In spite of the fact that the DFTB and MDTB models predict comparable $C_{2v}-I_h$ energy differences and barrier heights, their MEPs have differences, as well as similarities. The MDTB path results in a much narrower barrier. Analyzing the structures along the MDTB path, one can see that as the path moves away from the $I_h$ start configuration, the $\mbox{C}_{60}$ sphere distorts to become an ovoid before any SW rotation happens; the sharp energy rise that results in the barrier occurs only once chemical bonds break to allow for the dimer rotation to occur. In the case of  the DFTB model, however, dimer rotation begins much sooner, resulting in a wider barrier.  This is consistent with the fact that the MDTB model predicts much softer vibrational frequencies than the DFTB model for $\mbox{C}_{60}$. 

The saddle point configuration closest to the maximum of the MEP shown in Fig.~(\ref{fig:C60MEP}) has $C_2$ point-group symmetry. It is very similar
to the structure with the same symmetry reported by Bettinger {\em et al.\/}~\cite{Bettinger:1915p1573} on the basis of DFT calculations. The same authors reported a second, asymmetric, transition state involving a carbene intermediate. We do not find such a structure with either TB model used in this work. It is interesting that Walsh and Wales~\cite{Walsh:1998p1642} found a different saddle point configuration to the one we obtain, even though in principle they used the DFTB model. Their structure is asymmetric, with the rotating C-C bond highlighted in red in Fig.~(\ref{fig:C60MEP}) tilted towards one side of the open cage, forming a triangle with the under-coordinated atom at the vertex of the nearby hexagon (see Fig.~1 of their paper). We have also searched for this structure using the RFO~\cite{Banerjee:85,Baker:1886p1649} method, but have not been able to locate it. Starting the RFO search from a structure similar to theirs converges to the same symmetric $C_2$ structure that we find from our calculated MEP. The reason for this discrepancy remains unclear, but it is most likely due to the use of different parametrizations of the same TB model in their work and ours.  

\section{Conclusions}
\label{sec:conclusions}

We have presented a new algorithm for the search of minimum energy paths (MEPs) between two given stationary points on a potential energy surface, 
and demonstrated its use and robustness when applied to a number of low-dimensional toy models and to the Stone-Wales transformation in $C_{60}$. Our method falls in the class of algorithms known as {\em chain-of-states methods\/}, requiring only gradient information from the PES. But in contrast to well-known examples of other methods of this kind, such as the {\em Nudged-Elastic Band\/}, it does not rely on the introduction of additional force terms acting  on the states, nor does it rely on redistributing the states along the path, as in the {\em String\/} method, to converge to the MEP.  In spite of this, in all the test cases we have considered the method converges smoothly, retaining the images discretizing the path approximately evenly spaced over its length. Because the method uses an analytical representation of the path, it is possible, if desired, to refine the path discretization by inserting new images where required, or to scale the tangential acceleration components to ensure an even spacing, although we stress that in the test calculations we have reported above this was found to be unnecessary. The method has an appealing simplicity, making it easy to implement and combine with existing atomistic simulation codes. 

Our practical implementation of the scheme is robust and efficient. Nevertheless it is susceptible to improvement in a number of ways. In particular, the acceleration update scheme (step 3 of the method) is akin to a steepest-descent method. Previous experience in the NEB and String methods has shown that more sophisticated update schemes can significantly improve the rate of convergence towards the MEP, and we expect this to be the case here as well. We will explore this issue in future research. We also hope to demonstrate the usefulness of the present scheme in a wider set of case studies. 

\section*{Acknowledgements}
This work has been supported by the Spanish Research and Innovation Office through projects No. FIS2012-31713 and FIS2012-37549.
%\appendix

%\section{The modified LEPS potential}
%\label{app:LEPS}

%\section{The M\"{u}ller potential}
%\label{app:Muller}

%\bibliography{/Users/ehe/LaTex/papers/bibliography/books,/Users/ehe/LaTex/papers/bibliography/simulation,/Users/ehe/LaTex/papers/bibliography/MEP,/Users/ehe/LaTex/papers/bibliography/nanotubes,/Users/ehe/LaTex/papers/bibliography/electronicstructure,/Users/ehe/LaTex/papers/bibliography/clusters}

\begin{thebibliography}{29}
\expandafter\ifx\csname natexlab\endcsname\relax\def\natexlab#1{#1}\fi
\expandafter\ifx\csname bibnamefont\endcsname\relax
  \def\bibnamefont#1{#1}\fi
\expandafter\ifx\csname bibfnamefont\endcsname\relax
  \def\bibfnamefont#1{#1}\fi
\expandafter\ifx\csname citenamefont\endcsname\relax
  \def\citenamefont#1{#1}\fi
\expandafter\ifx\csname url\endcsname\relax
  \def\url#1{\texttt{#1}}\fi
\expandafter\ifx\csname urlprefix\endcsname\relax\def\urlprefix{URL }\fi
\providecommand{\bibinfo}[2]{#2}
\providecommand{\eprint}[2][]{\url{#2}}

\bibitem[{\citenamefont{J{\'o}nsson et~al.}(1998)\citenamefont{J{\'o}nsson,
  Mills, and Jacobsen}}]{Jonsson:1998p1578}
\bibinfo{author}{\bibfnamefont{H.}~\bibnamefont{J{\'o}nsson}},
  \bibinfo{author}{\bibfnamefont{G.}~\bibnamefont{Mills}}, \bibnamefont{and}
  \bibinfo{author}{\bibfnamefont{K.~W.} \bibnamefont{Jacobsen}},
 \emph{\bibinfo{booktitle}{Classical and Quantum Dynamics in Condensed Phase
  Simulations}}, edited by \bibinfo{editor}{\bibfnamefont{B.~J.}
  \bibnamefont{Berne}},
  \bibinfo{editor}{\bibfnamefont{G.}~\bibnamefont{Ciccotti}}, \bibnamefont{and}
  \bibinfo{editor}{\bibfnamefont{D.~F.} \bibnamefont{Cocker}}
  (\bibinfo{publisher}{World Scientific}, \bibinfo{year}{1998}),
  chap.~\bibinfo{chapter}{16}, p.~\bibinfo{pages}{385}. 

\bibitem[{\citenamefont{Chandler}(1998)}]{Chandler:1998}
\bibinfo{author}{\bibfnamefont{D.}~\bibnamefont{Chandler}}, in
  \emph{\bibinfo{booktitle}{Classical and Quantum Dynamics in Condensed Phase
  Simulations}}, edited by \bibinfo{editor}{\bibfnamefont{B.~J.}
  \bibnamefont{Berne}},
  \bibinfo{editor}{\bibfnamefont{G.}~\bibnamefont{Ciccotti}}, \bibnamefont{and}
  \bibinfo{editor}{\bibfnamefont{D.~F.} \bibnamefont{Cocker}}
  (\bibinfo{publisher}{World Scientific}, \bibinfo{year}{1998}),
  chap.~\bibinfo{chapter}{3}, p.~\bibinfo{pages}{51}.

\bibitem[{\citenamefont{Bolhuis et~al.}(2002)\citenamefont{Bolhuis, Chandler,
  Dellago, and Geissler}}]{Bolhuis:2002p1668}
\bibinfo{author}{\bibfnamefont{P.~G.} \bibnamefont{Bolhuis}},
  \bibinfo{author}{\bibfnamefont{D.}~\bibnamefont{Chandler}},
  \bibinfo{author}{\bibfnamefont{C.}~\bibnamefont{Dellago}}, \bibnamefont{and}
  \bibinfo{author}{\bibfnamefont{P.~L.} \bibnamefont{Geissler}},
  \bibinfo{journal}{Annu. Rev. Phys. Chem.} \textbf{\bibinfo{volume}{53}},
  \bibinfo{pages}{291} (\bibinfo{year}{2002}).

\bibitem[{\citenamefont{Dellago et~al.}(1998)\citenamefont{Dellago, Bolhuis,
  Csajka, and Chandler}}]{Dellago:1998p1669}
\bibinfo{author}{\bibfnamefont{C.}~\bibnamefont{Dellago}},
  \bibinfo{author}{\bibfnamefont{P.~G.} \bibnamefont{Bolhuis}},
  \bibinfo{author}{\bibfnamefont{F.~S.} \bibnamefont{Csajka}},
  \bibnamefont{and} \bibinfo{author}{\bibfnamefont{D.}~\bibnamefont{Chandler}},
  \bibinfo{journal}{J. Chem. Phys.}
  \textbf{\bibinfo{volume}{108}}, \bibinfo{pages}{1964} (\bibinfo{year}{1998}).

\bibitem[{\citenamefont{Sevick et~al.}(1993)\citenamefont{Sevick, Bell, and
  Theodorou}}]{Sevick:1993p1548}
\bibinfo{author}{\bibfnamefont{E.~M.} \bibnamefont{Sevick}},
  \bibinfo{author}{\bibfnamefont{A.~T.} \bibnamefont{Bell}}, \bibnamefont{and}
  \bibinfo{author}{\bibfnamefont{D.~N.} \bibnamefont{Theodorou}},
  \bibinfo{journal}{J. Chem. Phys.} \textbf{\bibinfo{volume}{98}},
  \bibinfo{pages}{3196} (\bibinfo{year}{1993}).

\bibitem[{\citenamefont{Wales}(2002)}]{Wales:2002p1616}
\bibinfo{author}{\bibfnamefont{D.~J.} \bibnamefont{Wales}},
  \bibinfo{journal}{Mol. Phys.} \textbf{\bibinfo{volume}{100}},
  \bibinfo{pages}{3285} (\bibinfo{year}{2002}).
  
\bibitem[{\citenamefont{Banerjee et~al.}(1985)\citenamefont{Banerjee, Adams,
  Simons, and Shepard}}]{Banerjee:85}
\bibinfo{author}{\bibfnamefont{A.}~\bibnamefont{Banerjee}},
  \bibinfo{author}{\bibfnamefont{N.}~\bibnamefont{Adams}},
  \bibinfo{author}{\bibfnamefont{J.}~\bibnamefont{Simons}}, \bibnamefont{and}
  \bibinfo{author}{\bibfnamefont{R.}~\bibnamefont{Shepard}},
  \bibinfo{journal}{J.  Phys. Chem.}
  \textbf{\bibinfo{volume}{89}}, \bibinfo{pages}{1} (\bibinfo{year}{1985}).

\bibitem[{\citenamefont{Baker}(1886)}]{Baker:1886p1649}
\bibinfo{author}{\bibfnamefont{J.}~\bibnamefont{Baker}},
  \bibinfo{journal}{J. Comput. Chem.}
  \textbf{\bibinfo{volume}{7}}, \bibinfo{pages}{385} (\bibinfo{year}{1886}).
  
  \bibitem[{\citenamefont{Munro and Wales}(1999)}]{Munro:1999p1643}
\bibinfo{author}{\bibfnamefont{L.~J.} \bibnamefont{Munro}} \bibnamefont{and}
  \bibinfo{author}{\bibfnamefont{D.~J.} \bibnamefont{Wales}},
  \bibinfo{journal}{Physical Review B}
  \textbf{\bibinfo{volume}{59}}, \bibinfo{pages}{3969} (\bibinfo{year}{1999}).
  
  \bibitem[{\citenamefont{Kumeda et~al.}(2001)\citenamefont{Kumeda, Wales, and
  Munro}}]{Kumeda:2001p1683}
\bibinfo{author}{\bibfnamefont{Y.}~\bibnamefont{Kumeda}},
  \bibinfo{author}{\bibfnamefont{D.~J.} \bibnamefont{Wales}}, \bibnamefont{and}
  \bibinfo{author}{\bibfnamefont{L.~J.} \bibnamefont{Munro}},
  \bibinfo{journal}{Chemical Physics Letters} \textbf{\bibinfo{volume}{341}},
  \bibinfo{pages}{185} (\bibinfo{year}{2001}).

\bibitem[{\citenamefont{Zeng et~al.}(2014)\citenamefont{Zeng, Xiao, and
  Henkelman}}]{Zeng:2014p1681}
\bibinfo{author}{\bibfnamefont{Y.}~\bibnamefont{Zeng}},
  \bibinfo{author}{\bibfnamefont{P.}~\bibnamefont{Xiao}}, \bibnamefont{and}
  \bibinfo{author}{\bibfnamefont{G.}~\bibnamefont{Henkelman}},
  \bibinfo{journal}{The Journal of Chemical Physics}
  \textbf{\bibinfo{volume}{140}}, \bibinfo{pages}{044115}
  (\bibinfo{year}{2014}).

\bibitem[{\citenamefont{Henkelman and J{\'o}nsson}(1999)}]{Henkelman:1999p1654}
\bibinfo{author}{\bibfnamefont{G.}~\bibnamefont{Henkelman}} \bibnamefont{and}
  \bibinfo{author}{\bibfnamefont{H.}~\bibnamefont{J{\'o}nsson}},
  \bibinfo{journal}{J. Chem. Phys.} \textbf{\bibinfo{volume}{111}},
  \bibinfo{pages}{7010} (\bibinfo{year}{1999}).

\bibitem[{\citenamefont{Page and McIver}(1988)}]{Page:1988p1650}
\bibinfo{author}{\bibfnamefont{M.}~\bibnamefont{Page}} \bibnamefont{and}
  \bibinfo{author}{\bibfnamefont{J.~W.} \bibnamefont{McIver}},
  \bibinfo{journal}{J. Chem. Phys.} \textbf{\bibinfo{volume}{88}},
  \bibinfo{pages}{922} (\bibinfo{year}{1988}).

\bibitem[{\citenamefont{Henkelman et~al.}(2000)\citenamefont{Henkelman,
  Uberuaga, and J{\'o}nsson}}]{Henkelman:2000p1539}
\bibinfo{author}{\bibfnamefont{G.}~\bibnamefont{Henkelman}},
  \bibinfo{author}{\bibfnamefont{B.~P.} \bibnamefont{Uberuaga}},
  \bibnamefont{and}
  \bibinfo{author}{\bibfnamefont{H.}~\bibnamefont{J{\'o}nsson}},
  \bibinfo{journal}{J. Chem. Phys.} \textbf{\bibinfo{volume}{113}},
  \bibinfo{pages}{9901} (\bibinfo{year}{2000}).

\bibitem[{\citenamefont{Henkelman and J{\'o}nsson}(2000)}]{Henkelman:2000p1655}
\bibinfo{author}{\bibfnamefont{G.}~\bibnamefont{Henkelman}} \bibnamefont{and}
  \bibinfo{author}{\bibfnamefont{H.}~\bibnamefont{J{\'o}nsson}},
  \bibinfo{journal}{J. Chem. Phys.} \textbf{\bibinfo{volume}{113}},
  \bibinfo{pages}{9978} (\bibinfo{year}{2000}).
  
 \bibitem[{\citenamefont{Trygubenko and Wales}(2004)}]{Trygubenko:2004p1678}
\bibinfo{author}{\bibfnamefont{S.~A.} \bibnamefont{Trygubenko}}
  \bibnamefont{and} \bibinfo{author}{\bibfnamefont{D.~J.} \bibnamefont{Wales}},
  \bibinfo{journal}{The Journal of Chemical Physics}
  \textbf{\bibinfo{volume}{120}}, \bibinfo{pages}{2082} (\bibinfo{year}{2004}).

\bibitem[{\citenamefont{E et~al.}(2002)\citenamefont{E, Ren, and
  Vanden-Eijnden}}]{E:2002p1547}
\bibinfo{author}{\bibfnamefont{W.}~\bibnamefont{E}},
  \bibinfo{author}{\bibfnamefont{W.}~\bibnamefont{Ren}}, \bibnamefont{and}
  \bibinfo{author}{\bibfnamefont{E.}~\bibnamefont{Vanden-Eijnden}},
  \bibinfo{journal}{Phys. Rev. B} \textbf{\bibinfo{volume}{66}},
  \bibinfo{pages}{052301} (\bibinfo{year}{2002}).

\bibitem[{\citenamefont{E et~al.}(2007)\citenamefont{E, Ren, and
  Vanden-Eijnden}}]{E:2007p1546}
\bibinfo{author}{\bibfnamefont{W.}~\bibnamefont{E}},
  \bibinfo{author}{\bibfnamefont{W.}~\bibnamefont{Ren}}, \bibnamefont{and}
  \bibinfo{author}{\bibfnamefont{E.}~\bibnamefont{Vanden-Eijnden}},
  \bibinfo{journal}{J. Chem. Phys.} \textbf{\bibinfo{volume}{126}},
  \bibinfo{pages}{164103} (\bibinfo{year}{2007}).
  
 \bibitem[{\citenamefont{Koslover and Wales}(2007)}]{Koslover:2007p1682}
\bibinfo{author}{\bibfnamefont{E.~F.} \bibnamefont{Koslover}} \bibnamefont{and}
  \bibinfo{author}{\bibfnamefont{D.~J.} \bibnamefont{Wales}},
  \bibinfo{journal}{The Journal of Chemical Physics}
  \textbf{\bibinfo{volume}{127}}, \bibinfo{pages}{134102}
  (\bibinfo{year}{2007}).

\bibitem[{\citenamefont{Sheppard et~al.}(2008)\citenamefont{Sheppard, Terrell,
  and Henkelman}}]{Sheppard:2008p1538}
\bibinfo{author}{\bibfnamefont{D.}~\bibnamefont{Sheppard}},
  \bibinfo{author}{\bibfnamefont{R.}~\bibnamefont{Terrell}}, \bibnamefont{and}
  \bibinfo{author}{\bibfnamefont{G.}~\bibnamefont{Henkelman}},
  \bibinfo{journal}{J. Chem. Phys.} \textbf{\bibinfo{volume}{128}},
  \bibinfo{pages}{134106} (\bibinfo{year}{2008}).
  
 \bibitem[]{code}{A simple Python script implementing this methodology in combination with the 
 LEPS and M\"{u}ller  potentials is available upon request from the corresponding author.}

\bibitem[{\citenamefont{Mueller and Brown}(1979)}]{Mueller:1979p1652}
\bibinfo{author}{\bibfnamefont{K.}~\bibnamefont{Mueller}} \bibnamefont{and}
  \bibinfo{author}{\bibfnamefont{L.~D.} \bibnamefont{Brown}},
  \bibinfo{journal}{Theoret. Chim. Acta} \textbf{\bibinfo{volume}{53}},
  \bibinfo{pages}{75} (\bibinfo{year}{1979}).

\bibitem[{\citenamefont{Stone and Wales}(1986)}]{stone86}
\bibinfo{author}{\bibfnamefont{A.~J.} \bibnamefont{Stone}} \bibnamefont{and}
  \bibinfo{author}{\bibfnamefont{D.~J.} \bibnamefont{Wales}},
  \bibinfo{journal}{Chem. Phys. Lett.} \textbf{\bibinfo{volume}{128}},
  \bibinfo{pages}{501} (\bibinfo{year}{1986}).

\bibitem[{\citenamefont{Kroto}(1987)}]{Kroto:1987p1670}
\bibinfo{author}{\bibfnamefont{H.~W.} \bibnamefont{Kroto}},
  \bibinfo{journal}{Nature} \textbf{\bibinfo{volume}{329}},
  \bibinfo{pages}{529} (\bibinfo{year}{1987}).

\bibitem[{\citenamefont{Hern\'{a}ndez et~al.}(2001)\citenamefont{Hern\'{a}ndez,
  Ordej\'{o}n, and Terrones}}]{hernandez:2001}
\bibinfo{author}{\bibfnamefont{E.~R.} \bibnamefont{Hern\'{a}ndez}},
  \bibinfo{author}{\bibfnamefont{P.}~\bibnamefont{Ordej\'{o}n}},
  \bibnamefont{and} \bibinfo{author}{\bibfnamefont{H.}~\bibnamefont{Terrones}},
  \bibinfo{journal}{Phys. Rev. B} \textbf{\bibinfo{volume}{63}},
  \bibinfo{pages}{193403} (\bibinfo{year}{2001}).

\bibitem[{\citenamefont{Nardelli
  et~al.}(1998{\natexlab{a}})\citenamefont{Nardelli, Yakobson, and
  Bernholc}}]{buongiorno98a}
\bibinfo{author}{\bibfnamefont{M.~B.} \bibnamefont{Nardelli}},
  \bibinfo{author}{\bibfnamefont{B.~I.} \bibnamefont{Yakobson}},
  \bibnamefont{and} \bibinfo{author}{\bibfnamefont{J.}~\bibnamefont{Bernholc}},
  \bibinfo{journal}{Phys. Rev. B} \textbf{\bibinfo{volume}{57}},
  \bibinfo{pages}{R4277} (\bibinfo{year}{1998}{\natexlab{a}}).

\bibitem[{\citenamefont{Nardelli
  et~al.}(1998{\natexlab{b}})\citenamefont{Nardelli, Yakobson, and
  Bernholc}}]{buongiorno98b}
\bibinfo{author}{\bibfnamefont{M.~B.} \bibnamefont{Nardelli}},
  \bibinfo{author}{\bibfnamefont{B.~I.} \bibnamefont{Yakobson}},
  \bibnamefont{and} \bibinfo{author}{\bibfnamefont{J.}~\bibnamefont{Bernholc}},
  \bibinfo{journal}{Phys. Rev. Lett.} \textbf{\bibinfo{volume}{81}},
  \bibinfo{pages}{4656} (\bibinfo{year}{1998}{\natexlab{b}}).
  
 \bibitem[{\citenamefont{Kumeda and Wales}(2003)}]{Kumeda:2003p1684}
\bibinfo{author}{\bibfnamefont{Y.}~\bibnamefont{Kumeda}} \bibnamefont{and}
  \bibinfo{author}{\bibfnamefont{D.~J.} \bibnamefont{Wales}},
  \bibinfo{journal}{Chemical Physics Letters} \textbf{\bibinfo{volume}{374}},
  \bibinfo{pages}{125} (\bibinfo{year}{2003}).

\bibitem[{\citenamefont{Bettinger et~al.}(2003)\citenamefont{Bettinger, Yakobson, and
  Scuseria}}]{Bettinger:1915p1573}
\bibinfo{author}{\bibfnamefont{H.~F.} \bibnamefont{Bettinger}},
  \bibinfo{author}{\bibfnamefont{B.}~\bibnamefont{Yakobson}}, \bibnamefont{and}
  \bibinfo{author}{\bibfnamefont{G.~E.} \bibnamefont{Scuseria}},
  \bibinfo{journal}{J. Am. Chem. Soc.}
  \textbf{\bibinfo{volume}{125}}, \bibinfo{pages}{5572} (\bibinfo{year}{2003}).

\bibitem[{\citenamefont{Tersoff}(1988)}]{tersoff88}
\bibinfo{author}{\bibfnamefont{J.}~\bibnamefont{Tersoff}},
  \bibinfo{journal}{Phys. Rev. B} \textbf{\bibinfo{volume}{37}},
  \bibinfo{pages}{6991} (\bibinfo{year}{1988}).

\bibitem[{\citenamefont{Xu et~al.}(1992)\citenamefont{Xu, Wang, Chan, and
  Ho}}]{Xu:1992p1639}
\bibinfo{author}{\bibfnamefont{C.}~\bibnamefont{Xu}},
  \bibinfo{author}{\bibfnamefont{C.}~\bibnamefont{Wang}},
  \bibinfo{author}{\bibfnamefont{C.}~\bibnamefont{Chan}}, \bibnamefont{and}
  \bibinfo{author}{\bibfnamefont{K.}~\bibnamefont{Ho}},
  \bibinfo{journal}{J. Phys.: Condens. Matter}
  \textbf{\bibinfo{volume}{4}}, \bibinfo{pages}{6047} (\bibinfo{year}{1992}).

\bibitem[{\citenamefont{Porezag et~al.}(1995)\citenamefont{Porezag, Frauenheim,
  Koehler, Seifert, and Kaschner}}]{DFTB}
\bibinfo{author}{\bibfnamefont{D.}~\bibnamefont{Porezag}},
  \bibinfo{author}{\bibfnamefont{T.}~\bibnamefont{Frauenheim}},
  \bibinfo{author}{\bibfnamefont{T.}~\bibnamefont{Koehler}},
  \bibinfo{author}{\bibfnamefont{G.}~\bibnamefont{Seifert}}, \bibnamefont{and}
  \bibinfo{author}{\bibfnamefont{R.}~\bibnamefont{Kaschner}},
  \bibinfo{journal}{Phys. Rev. B} \textbf{\bibinfo{volume}{51}},
  \bibinfo{pages}{12947} (\bibinfo{year}{1995}).

\bibitem[{\citenamefont{Rurali and Hern\'andez}(2003)}]{trocadero}
\bibinfo{author}{\bibfnamefont{R.}~\bibnamefont{Rurali}} \bibnamefont{and}
  \bibinfo{author}{\bibfnamefont{E.}~\bibnamefont{Hern\'andez}},
  \bibinfo{journal}{Comput. Mat. Sci.} \textbf{\bibinfo{volume}{28}},
  \bibinfo{pages}{85} (\bibinfo{year}{2003}).

\bibitem[{\citenamefont{Marcos et~al.}(1997)\citenamefont{Marcos, L{\'o}pez,
  Rubio, and Alonso}}]{Marcos:1997p1661}
\bibinfo{author}{\bibfnamefont{P.~A.} \bibnamefont{Marcos}},
  \bibinfo{author}{\bibfnamefont{M.~J.} \bibnamefont{L{\'o}pez}},
  \bibinfo{author}{\bibfnamefont{A.}~\bibnamefont{Rubio}}, \bibnamefont{and}
  \bibinfo{author}{\bibfnamefont{J.~A.} \bibnamefont{Alonso}},
  \bibinfo{journal}{Chem. Phys. Lett.} \textbf{\bibinfo{volume}{273}},
  \bibinfo{pages}{367} (\bibinfo{year}{1997}).

\bibitem[{\citenamefont{Goringe et~al.}(1997)\citenamefont{Goringe, Bowler, and
  Hern\'{a}ndez}}]{goringe97}
\bibinfo{author}{\bibfnamefont{C.~M.} \bibnamefont{Goringe}},
  \bibinfo{author}{\bibfnamefont{D.~R.} \bibnamefont{Bowler}},
  \bibnamefont{and} \bibinfo{author}{\bibfnamefont{E.~R.}
  \bibnamefont{Hern\'{a}ndez}}, \bibinfo{journal}{Rep. Prog. Phys.}
  \textbf{\bibinfo{volume}{60}}, \bibinfo{pages}{1447} (\bibinfo{year}{1997}).

\bibitem[{\citenamefont{Walsh and Wales}(1998)}]{Walsh:1998p1642}
\bibinfo{author}{\bibfnamefont{T.~R.} \bibnamefont{Walsh}} \bibnamefont{and}
  \bibinfo{author}{\bibfnamefont{D.~J.} \bibnamefont{Wales}},
  \bibinfo{journal}{J. Chem. Phys.}
  \textbf{\bibinfo{volume}{109}}, \bibinfo{pages}{6691} (\bibinfo{year}{1998}).

\end{thebibliography}

\end{document}